\documentclass[pre,twocolumn,10pt,aps,longbibliography,nofootinbib]{revtex4-1}

 \usepackage{verbatim}
 \usepackage{amsmath}
 \usepackage{amssymb}
 \usepackage{dsfont}
 \usepackage{latexsym}
 \usepackage{amsfonts}
 \usepackage{epsfig}
 \usepackage{epstopdf}
 \usepackage{color}
 \definecolor{darkblue}{rgb}{0,0,.5}
 \usepackage[linktocpage, colorlinks=true, linkcolor=darkblue, citecolor=darkblue]{hyperref}
 \usepackage[all]{hypcap}

\newcommand{\C}[1]{{\cal{#1}}}
\newcommand{\bb}[1]{\textbf{#1}}

\newcommand{\lr}[1]{{\left\langle {#1}\right\rangle}}

\begin{document}

\title{Stochastic thermodynamics in the strong coupling regime: \\ 
An unambiguous approach based on coarse-graining}

\author{Philipp Strasberg}
\email{philipp.strasberg@uni.lu}
\author{Massimiliano Esposito}
\affiliation{Complex Systems and Statistical Mechanics, Physics and Materials Science, University of Luxembourg, L-1511 Luxembourg, Luxembourg}

\date{\today}

\begin{abstract}
 We consider a classical and possibly driven composite system $X \otimes Y$ weakly coupled to a Markovian 
 thermal reservoir $R$ so that an unambiguous stochastic thermodynamics ensues for $X \otimes Y$. 
 This setup can be equivalently seen as a system $X$ strongly coupled to a non-Markovian reservoir $Y \otimes R$. 
 We demonstrate that only in the limit where the dynamics of $Y$ is much faster than $X$, 
 our unambiguous expressions for thermodynamic quantities such as heat, entropy or internal energy,
 are equivalent to the strong coupling expressions recently obtained in the literature using the 
 Hamiltonian of mean force. By doing so, we also significantly extend these results by formulating 
 them at the level of instantaneous rates and by allowing for time-dependent couplings between 
 $X$ and its environment. Away from the limit where $Y$ evolves much faster than $X$, previous approaches fail to 
 reproduce the correct results from the original unambiguous formulation, as we illustrate numerically for an 
 underdamped Brownian particle coupled strongly to a non-Markovian reservoir. 
\end{abstract}

\maketitle

\section{Introduction}

Establishing the laws of thermodynamics for a given setup is not only beneficial for practical purposes, but also 
provides an important consistency check for the validity of the model and provides much deeper insights into the 
structure of the problem. 
Yet, establishing these laws for small-scale systems away from the well-established weak-coupling and Markovian limit 
can be very challenging. 

This paper focuses on the case of a small, driven classical system in strong contact with a single 
environment. This case has attracted a lot of attention recently and was mostly tackled by introducing a 
\emph{Hamiltonian of mean force} (HMF), classically~\cite{JarzynskiJSM2004, GelinThossPRE2009, SeifertPRL2016, 
TalknerHaenggiPRE2016, PhilbinAndersJPA2016, JarzynskiPRX2017, MillerAndersArXiv2017} as well as 
quantum-mechanically~\cite{GelinThossPRE2009, CampisiTalknerHaenggiPRL2009, CampisiTalknerHaenggiJPA2009, 
HiltEtAlPRE2011, PucciEspositoPelitiJSM2013, PhilbinAndersJPA2016}. However, the question of what exactly are the 
correct definitions for heat, internal energy and other quantities causes already controversies at the classical 
level~\cite{GelinThossPRE2009, SeifertPRL2016, TalknerHaenggiPRE2016, JarzynskiPRX2017}. 

We present an enlightening perspective on this problem by considering two coupled systems $X\otimes Y$ which are in 
weak contact with a large thermal reservoir $R$ and obey standard stochastic thermodynamics. By realizing that 
the system $X$ can be strongly coupled to the system $Y$, we see that the situation is \emph{equivalent} to a system $X$ 
in strong contact with an environment $E = Y\otimes R$. In fact, for many relevant scenarios it makes sense that the 
system $X$ only couples strongly to a subpart $Y\subset E$ of the environment, but not to each degree of freedom of $E$. 
This picture is also supported by our example at the end of the paper. 

The benefit of our approach is that we start from a well-defined thermodynamics with unambiguous definitions and we 
can then compare under which conditions previous approaches based on the HMF coincide with them. Our framework can be 
seen as an applications of the laws of thermodynamics under coarse graining as detailed in 
Ref.~\cite{EspositoPRE2012}, also see Refs.~\cite{SeifertEPJE2011, BoCelaniJSM2014}. 

\emph{Outline: } We start by presenting the thermodynamic description of the combined system $X\otimes Y$ in contact 
with $R$ in Sec.~\ref{sec setup}. We show what changes if we coarse-grain $Y$ and consider the important limit where 
$Y$ evolves much faster than $X$ such that it can be adiabatically eliminated and a closed thermodynamic description for 
$X$ alone emerges. In Sec.~\ref{sec connection HMF} we turn the situation around and start from a description of $X$ 
coupled to $E$ and derive an exact inequality. Based on this we recapitulate the thermodynamics previously 
established using the HMF and we show that we are able to rederive and greatly extend these results if $Y$ evolves much 
faster than $X$. Beyond that we explicitly quantify the difference in the two proposed definitions of entropy 
production and we use an example in Sec.~\ref{sec example} to illustrate generic features in the 
thermodynamics of strongly coupled systems demonstrating that knowledge of the HMF alone does not suffice to reproduce 
the original thermodynamics. 
It also provides a strategy to identify a system $Y$ if the initial setup is described at the level of $X\otimes E$. 
Sec.~\ref{sec summary} summarizes the main implications of our findings.

\section{Setup}
\label{sec setup}

\subsection{Basic quantities}

\begin{figure}
 \centering\includegraphics[width=0.33\textwidth,clip=true]{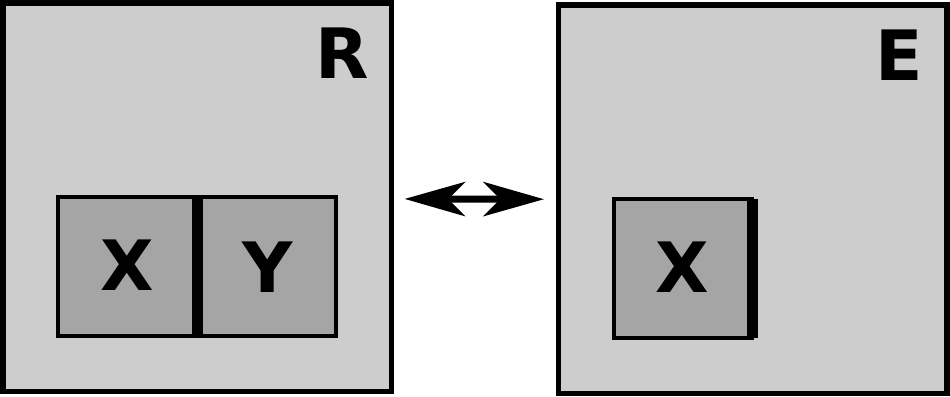}
 \label{fig setup} 
 \caption{Sketch of the setup: two systems $X$ and $Y$ weakly interact with a large thermal reservoir $R$. Equivalently, 
 the system $X$ interacts strongly with a composite environment $E \equiv Y\otimes R$. }
\end{figure}

A sketch of the setup is shown in Fig.~\ref{fig setup}. Two systems $X$ and $Y$ interact with each other and their 
joint Hamiltonian is assumed to be of the form 
\begin{equation}
 E_{xy}(\lambda_t) = E_x(\lambda_t) + V_{xy}(\lambda_t) + E_y.
\end{equation}
Here, the energy $E_x(\lambda_t)$ of system $X$ as well as the interaction energy $V_{xy}(\lambda_t)$ can be 
time-dependent due to some externally controlled parameters $\lambda_t$. The energy $E_y$ of system $Y$ is assumed to be 
time-independent. 

Since the joint system $X\otimes Y$ is weakly coupled to a large thermal reservoir at inverse temperature $\beta$, 
we assume that its dynamics can be modeled by a Markovian master equation (ME) of the form
\begin{equation}\label{eq ME}
 d_t p_{xy}(t) = \sum_{x',y'} R_{xy,x'y'}(\lambda_t) p_{x'y'}(t).
\end{equation}
Here, $p_{xy}(t)$ denotes the probability to find the system in state $xy$ at time $t$. The rate matrix obeys 
$\sum_{x,y} R_{xy,x'y'}(\lambda_t) = 0$ which ensures conservation of probability [$\sum_{x,y} p_{xy}(t) = 1$]. 
Furthermore, we assume local detailed balance 
\begin{equation}
 \frac{R_{xy,x'y'}(\lambda_t)}{R_{x'y',xy}(\lambda_t)} = e^{-\beta[E_{xy}(\lambda_t) - E_{x'y'}(\lambda_t)]},
\end{equation}
which allows a physical interpretation of the ME and especially a consistent thermodynamic description. 
The rest in this section then follows standard stochastic thermodynamics~\cite{SchnakenbergRMP1976, SeifertRPP2012, 
VandenBroeckEspositoPhysA2015}.

For this purpose we introduce the following quantities: 
\begin{align}
 U_{XY}(t)	&\equiv	\lr{E_{xy}(\lambda_t)} ~~~ (\text{internal energy}),	\\
 \dot W(t)	&\equiv \lr{d_t E_{xy}(\lambda_t)} ~~~ (\text{work rate}),	\label{eq work}	\\
 \dot Q(t)	&\equiv	\sum_{x,y} E_{xy}(\lambda_t) d_tp_{xy}(t) ~~~ (\text{heat rate}),	\label{eq heat}	\\
 S_{XY}(t)	&\equiv \lr{-\ln p_{xy}(t)} ~~~ (\text{entropy}).
\end{align}
Here, we have denoted the ensemble average with respect to any solution $p_{xy}(t)$ of the ME above by 
$\lr{f_{xy}(t)} \equiv \sum_{x,y} p_{xy}(t) f_{xy}(t)$. Furthermore, the thermodynamic entropy $S$ coincides in 
the weak coupling regime with the definition of Shannon entropy. 

Based on the definitions above, it is straightforward to derive the first law: 
\begin{equation}
 d_t U_{XY}(t) = \dot W(t) + \dot Q(t)
\end{equation}
(we define heat and work positive if they increase the energy of $X\otimes Y$). Furthermore, the second law states that 
the overall entropy production rate is postive: 
\begin{equation}
 \dot\Sigma(t) \equiv d_t S_{XY}(t) - \beta\dot Q(t) \ge 0.
\end{equation}
Its positity can be proven by noting the identity 
\begin{align}
 &	\dot \Sigma(t)	\\
 &=	\sum_{x,x',y,y'} R_{xy,x'y'}(\lambda_t) p_{x'y'}(t)\ln\frac{R_{xy,x'y'}(\lambda_t) p_{x'y'}(t)}{R_{x'y',xy}(\lambda_t) p_{xy}(t)}	\nonumber
\end{align}
and using $-\ln x \ge 1-x$. Another useful identity is 
\begin{equation}\label{eq entropy production alternative}
 \dot \Sigma(t) = -\partial_t|_{\lambda(t)} D[p_{xy}(t)\|p_{xy}^\text{eq}(\lambda_t)] \ge 0,
\end{equation}
where the partial derivative $\partial_t|_{\lambda(t)}$ indicates that the change of 
$D[p_{xy}(t)\|p_{xy}^\text{eq}(\lambda_t)]$ is evaluated at fixed $\lambda_t$. 
Here, we introduced the equilibrium (Gibbs, thermal) state 
\begin{equation}
 p_{xy}^\text{eq}(\lambda_t) \equiv \frac{e^{-\beta E_{xy}(\lambda_t)}}{\C Z_{XY}(\lambda_t)},
\end{equation}
which depends parametrically on time and $\C Z_{XY}(\lambda_t)$ denotes the equilibrium partition function. 
Also the concept of relative entropy, 
\begin{equation}
 D[p_x\|q_x] \equiv \sum_x p_x(\ln p_x - \ln q_x) \ge 0,
\end{equation}
which is always positive for any two probability distributions $p_x$ and $q_x$, will be used later on. 

Finally, let us introduce the concept of a non-equilibrium free energy $F_{XY}(t) = U_{XY}(t) - \beta^{-1} S_{XY}(t)$, 
which is defined for any state $p_{xy}(t)$. Using this, we can reformulate the second law as 
\begin{equation}\label{eq entropy production alternative 2}
 \dot \Sigma(t) = \beta[\dot W(t) - d_t F_{XY}(t)] \ge 0.
\end{equation}
Whenever a system $\alpha$ (where $\alpha$ could stand for $XY,X,E,\dots$ depending on the situation) 
is at equilibrium, it is useful to note the relations 
\begin{align}
 \C F_{\alpha}(\lambda_t)	&=	-\beta^{-1}\ln\C Z_{\alpha}(\lambda_t),	\\
 \C U_{\alpha}(\lambda_t)	&=	\partial_\beta[\beta\C F_{\alpha}(\lambda_t)],	\label{eq def U equilibrium}	\\
 \C S_{\alpha}(\lambda_t)	&=	\beta^2\partial_\beta \C F_{\alpha}(\lambda_t)	\label{eq def S equilibrium}
\end{align}
for the equilibrium free energy, internal energy and entropy.
Note that we use calligraphic letters $\C F, \C Z, \C U, \C S$ to denote thermodynamic quantities at equilibrium. 

Below, to keep a compact notation, we will often omit the dependence on $\lambda_t$ in the notation.

\subsection{Coarse-graining}

We now shift our attention to system $X$ alone and mostly follow Ref.~\cite{EspositoPRE2012} for the rest of this section. 
For this purpose we split the joint probability into a conditional and marginal probability as 
\begin{equation}
 p_{xy}(t) = p_{y|x}(t)p_x(t)
\end{equation}
with $p_x(t) = \sum_y p_{xy}(t)$ and $\sum_y p_{y|x}(t) = 1$. It is not hard to deduce that $p_x(t)$ evolves 
according to the ME 
\begin{equation}\label{eq ME effective X}
 d_t p_x(t) = \sum_{x'} R_{x,x'} p_{x'}(t)
\end{equation}
where the new effective rate matrix is given by 
\begin{equation}
 R_{x,x'} = R_{x,x'}(\lambda_t,t) = \sum_{y,y'} R_{xy,x'y'}(\lambda_t) p_{y'|x'}(t).
\end{equation}
In general, it depends explicitly on time due to the time-dependence of $p_{y'|x'}(t)$ and 
solving~(\ref{eq ME effective X}) is equally hard as solving the original ME unless further assumptions are made. 

Nevertheless, there is an apparent second law related to the reduced dynamics of $X$~\cite{EspositoPRE2012}: 
\begin{equation}
 \dot \Sigma^{(1)}(t) = \sum_{x,x'} R_{x,x'} p_{x'}(t) \ln\frac{R_{x,x'} p_{x'}(t)}{R_{x',x} p_{x}(t)} \ge 0,
\end{equation}
which can be rewritten as 
\begin{equation}
 \dot \Sigma^{(1)}(t) = d_t S_{XY}(t) - \beta \dot Q^{(1)}(t) \ge 0.
\end{equation}
Here, we introduced the apparent heat flow 
\begin{equation}\label{eq heat fictitious}
 \dot Q^{(1)}(t) \equiv -\frac{1}{\beta}\sum_{x,x'} R_{x,x'} p_{x'}(t) \ln\frac{R_{x,x'}}{R_{x',x}} + \frac{1}{\beta}d_t S_{Y|X}(t)
\end{equation}
and $S_{Y|X}$ denotes the conditional Shannon entropy 
\begin{equation}
 S_{Y|X}(t) = - \sum_x p_x(t)\sum_y p_{y|x}(t)\ln p_{y|x}(t),
\end{equation}
which fulfills $S_{XY} = S_X + S_{Y|X}$ with $S_X \equiv -\sum_x p_x(t)\ln p_x(t)$. 
Unfortunately, at this general level there is no relation between $\dot Q^{(1)}$ and the 
real heat flow $\dot Q$ making it hard to establish a local version of the first law. Furthermore, note that 
$\dot \Sigma^{(1)}$ always underestimates the true entropy production 
${\dot \Sigma \ge \dot \Sigma^{(1)} \ge 0}$.

\subsection{Time-scale separation}
\label{sec time-scale separation}

There is an important limit, in which $Y$ evolves much faster than $X$ and can be adiabatically eliminated. 
We will refer to this as time-scale separation (TSS). Within TSS one assumes that 
\begin{equation}\label{eq TSS limit}
 R_{xy,xy'} \gg R_{xy,x'y'}
\end{equation}
for $x\neq x'$. To get a simple description we also assume that for each given $x$ all states $y$ are connected, i.e., 
$R_{xy,xy'} \neq 0$ for all $y,y'$. 

Under these conditions one can show that the conditional probabilities $p_{y|x}(t)$ equilibrate and can be written 
as\footnote{Of course, there are also alternative parametrizations possible. For instance, 
$\bar p_{y|x} = e^{-\beta[E_y + V_{xy}(\lambda_t) - F'_{Y|x}(\lambda_t)]}$ with 
$F'_{Y|x}(\lambda_t) = F_{Y|x}(\lambda_t) - E_x(\lambda_t)$. This does not affect the resulting thermodynamics at the end. } 
\begin{equation}\label{eq cond equilibrium Y}
 \bar p_{y|x} = \bar p_{y|x}(\lambda_t) = e^{-\beta(E_{xy} - F_{Y|x})}.
\end{equation}
Normalization is ensured by choosing 
\begin{align}\label{eq free energy Y TSS}
 F_{Y|x}	&\equiv	-\beta^{-1} \ln\sum_y e^{-\beta E_{xy}(\lambda_t)}	\\
		&=	E_x - \beta^{-1}\ln\lr{e^{-\beta V_{xy}}}_Y^\text{eq} + \C F_Y,	\nonumber
\end{align}
where $\lr{\dots}_Y^\text{eq}$ denotes the ensemble average with respect to $p_y^\text{eq} = e^{-\beta E_y}/\C Z_Y$. 
Note that $F_{Y|x} = F_{Y|x}(\lambda_t)$ depends parametrically on time. In contrast, the equilibrium free energy 
$\C F_Y$ has no time-dependence. Although it appears in the definition of $F_{Y|x}$, we remark that the reduced state 
of $Y$ is not given by the equilibrium state $p_y^\text{eq}$. 

Within TSS we denote the rate matrix $R_{x,x'}$ by $\bar R_{x,x'} = \bar R_{x,x'}(\lambda_t)$, which now depends only 
parametrically on time and greatly simplifies the solution of Eq.~(\ref{eq ME effective X}). 
Furthermore, it fulfills an effective local detailed balance relation of the form 
\begin{equation}\label{eq local detailed balance TSS}
 \frac{\bar R_{x,x'}}{\bar R_{x',x}} = e^{-\beta(F_{Y|x} - F_{Y|x'})},
\end{equation}
which makes the meaning of $F_{Y|x}$ as a free energy landscape for system $X$ transparent. Using this, 
we can express the apparent heat flow~(\ref{eq heat fictitious}) as 
\begin{equation}\label{eq heat fictitious TSS}
 \dot Q^{(1)}(t) = \sum_{x,x'} F_{Y|x} \bar R_{x,x'}p_{x'}(t) + \frac{1}{\beta}d_t S_{Y|X}(t),
\end{equation}
which now coincides with the real heat flow: 
\begin{equation}\label{eq heat TSS}
 \dot Q(t) = \dot Q^{(1)}(t).
\end{equation}
To prove Eq.~(\ref{eq heat TSS}) it is useful to note that 
\begin{equation}
 \begin{split}
  \frac{1}{\beta} d_t S_{Y|X}(t)	=&	-\frac{1}{\beta}\sum_{x,y} [d_t p_x(t)\bar p_{y|x}(\lambda_t)] \ln \bar p_{y|x}(\lambda_t)	\\
					=&	\sum_{x,y} (E_{xy} - F_{Y|x}) [d_t p_x(t)\bar p_{y|x}(\lambda_t)],
 \end{split}
\end{equation}
where we used~(\ref{eq cond equilibrium Y}). Then, after realizing that ${\sum_y \bar p_{y|x}(\lambda_t) = 1}$ and 
$\sum_y d_t\bar p_{y|x}(\lambda_t) = 0$, we get 
\begin{equation}
 \begin{split}
  & \frac{1}{\beta} d_t S_{Y|X}(t) 	\\
  &=	\sum_{x,y} E_{xy} d_t [\bar p_{y|x}(\lambda_t) p_{x}(t)] + \sum_{x} F_{Y|x}  d_t p_{x}(t).
 \end{split}
\end{equation}
Plugging this result into Eq.~(\ref{eq heat fictitious TSS}), we finally obtain 
\begin{equation}
 \dot Q^{(1)}(t) = \sum_{x,y} E_{xy} d_t [\bar p_{y|x}(\lambda_t) p_{x}(t)],
\end{equation}
which equals our original definition~(\ref{eq heat}) within TSS. 

Furthermore, it makes sense to rewrite the internal energy as $U_{XY}(t) = \sum_x U_{XY|x} p_x(t)$
where we introduced the average internal energy \emph{conditioned} on the state $x$: 
\begin{equation}
 U_{XY|x} = U_{XY|x}(\lambda_t) \equiv \sum_y E_{xy}\bar p_{y|x}.
\end{equation}
Formally, the first law remains the same as before 
\begin{equation}
 d_t U_{XY}(t) = \dot W(t) + \dot Q(t) = \dot W(t) + \dot Q^{(1)}(t).	\label{eq 1st law TSS}
\end{equation}
In contrast to the general case, however, the time-dependence of all quantities comes only from the dynamical 
time-dependence of the system $X$ alone and the parametric dependence on $\lambda_t$. The same observation holds true for 
the second law of thermodynamics which can be expressed as 
\begin{equation}\label{eq 2nd law TSS}
 \dot \Sigma(t) = d_t S_{XY}(t) - \beta\dot Q^{(1)}(t) \ge 0.
\end{equation}
Thus, within TSS we have indeed $\dot \Sigma = \dot \Sigma^{(1)}$. 
For later purposes it will be also convenient to note the following two identities 
\begin{align}
 U_{XY|x}	&=	\partial_\beta(\beta F_{Y|x}),	\label{eq cond U TSS}	\\
 S_{Y|x}	&=	-\sum_y \bar p_{y|x}\ln \bar p_{y|x} = \beta^2\partial_\beta F_{Y|x},	\label{eq cond S TSS}
\end{align}
which look remarkably similar to Eqs.~(\ref{eq def U equilibrium}) and~(\ref{eq def S equilibrium}). 
Proving them follows from straightforward though tidious algebraic manipulations, which we will not display here. 

Finally, we briefly mention how to extend the results above to the stochastic level following the well-established 
procedure~\cite{SeifertRPP2012, VandenBroeckEspositoPhysA2015}. This will also underline the fact that any information 
about $Y$ enters only statically in the description. If the system starts at time $t_0$ in state $x_0$, 
jumps at time $t_1 > t_0$ to $x_1$ and stays in that state until it jumps at time $t_2 > t_1$ to $x_2$, etc., 
we denote this trajectory by $\bb x_t \equiv (x_0,t_0;x_1,t_1;\dots)$. 
Then, the fluctuating internal energy at each instant $t$ is given by 
\begin{equation}
 U_{XY|x_t} = \sum_y E_{x_t y} \bar p_{y|x_t}.
\end{equation}
The work along the trajectory $\bb x_t$ becomes 
\begin{equation}
 W[\bb x_t] = \int_{t_0}^t ds \dot\lambda_s\sum_y \bar p_{y|\bb x_s}\partial_{\lambda_s} E_{\bb x_s y}
\end{equation}
and the stochastic entropy is defined as 
\begin{equation}
 S_{XY}[x_t] = -\ln p_{x_t}(t) - \sum_y p_{y|x_t}(\lambda_t) \ln p_{y|x_t}(\lambda_t).
\end{equation}
Finally, the heat can be decomposed as 
\begin{align}
 &	\beta Q[\bb x_t] = \sum_j\ln\frac{\bar R_{x_j,x_{j+1}}}{\bar R_{x_{j+1},x_j}}	\\
 &-	\sum_y \bar p_{y|x_t}(\lambda_t) \ln \bar p_{y|x_t}(\lambda_t) + \sum_y \bar p_{y|x_0}(\lambda_0) \ln \bar p_{y|x_0}(\lambda_0)	\nonumber
\end{align}
where the sum indexed by $j$ runs over all jumps which have happened from $t_0$ to $t$. 
Since $p_x(t)$ obeys a Markovian ME with rates that fulfill the local detailed balance 
relation~(\ref{eq local detailed balance TSS}), it is clear that the integral and detailed fluctuation theorems 
are also obeyed, e.g., 
\begin{equation}
 \lr{\lr{e^{-\Sigma[\bb x_t]}}} = 1
\end{equation}
where $\Sigma[\bb x_t] = S_{XY}[x_t] - S_{XY}[x_0] - \beta Q[\bb x_t]$ and $\lr{\lr{\dots}}$ denotes an ensemble 
average over all trajectries $\bb x_t$.

\section{The Hamiltonian of mean force}
\label{sec connection HMF}

\subsection{Exact identities}
\label{sec HMF exact}

In this section we turn the situation around and consider a system $X$ in contact with an environment $E$ as shown on the 
right hand side of Fig.~\ref{fig setup} and we only use in Sec.~\ref{sec HMF reduced} the decomposition $E = Y\otimes R$. 
We assume that the combined system $X\otimes E$ is isolated and obeys an Hamiltonian dynamics with an Hamiltonian
\begin{equation}
E_{xe}(\lambda_t) = E_x(\lambda_t) + V_{xe}(\lambda_t) + E_e , \label{Tot_Hamiltonian}
\end{equation}
where $e$ denotes a microstate of the environment. 
Note that we will in general denote thermodynamic quantities in this section by a ``tilde'' to distinguish them from 
previously introduced quantities. Their relation will be clarified in Sec.~\ref{sec HMF reduced}. 

As in Ref.~\cite{SeifertPRL2016, TalknerHaenggiPRE2016}, we assume that the initial state of $X\otimes E$ reads 
\begin{equation}\label{eq initial state}
 p_{xe}(0) = p_x(0) \bar p_{e|x}(\lambda_0) = p_x(0) e^{-\beta(E_{xe} - F_{E|x})},
\end{equation}
where $p_x(0)$ is an arbitrary initial system state and $\bar p_{e|x}(\lambda_0)$ denotes the equilibrium state of $E$ 
conditioned on a microstate $x$ of the system. Clearly, $F_{E|x}$ is defined as in Eq.~(\ref{eq free energy Y TSS}) 
with $Y$ replaced by $E$. The state $\bar p_{e|x}$ can be more elegantly expressed by introducing the HMF, 
\begin{equation}\label{eq HMF standard definition}
 E_x^*(\lambda_t) \equiv E_x(\lambda_t) - \frac{1}{\beta}\ln\lr{e^{-\beta V_{xe}(\lambda_t)}}_E^\text{eq},
\end{equation}
which has been successfully used for a long time in thermostatics~\cite{KirkwoodJCP1935, RouxSimonsonBC1999}. 
Using this, we find 
\begin{equation}
 \bar p_{e|x}(\lambda_t) = \frac{e^{-\beta[E_{xe}(\lambda_t) - E_x^*(\lambda_t)]}}{\C Z_E}
\end{equation}
and also the important relation 
\begin{equation}\label{eq HMF}
 E_x^* = F_{E|x} - \C F_E.
\end{equation}

Given the initial state~(\ref{eq initial state}), we follow Ref.~\cite{EspositoLindenbergVandenBroeckNJP2010} 
and define an entropy production 
\begin{equation}\label{eq exact result}
 \tilde\Sigma(t) \equiv D[p_{xe}(t)\|p_{x}(t)\bar p_{e|x}(\lambda_t)] \ge 0,
\end{equation}
which measures the deviation of the true state $p_{xe}(t)$ from an idealized reference state 
$p_{x}(t)\bar p_{e|x}(\lambda_t)$. Note that definition~(\ref{eq exact result}) differs from 
Ref.~\cite{EspositoLindenbergVandenBroeckNJP2010} only in the choice of the reference state 
We discuss in Appendix~\ref{sec Appendix NJP} how both are related. 
We will now show that Eq.~(\ref{eq exact result}) coincides with the definition used in 
Ref.~\cite{SeifertPRL2016} as was independently and simultaneously noted in Ref.~\cite{MillerAndersArXiv2017}.

It is convenient to rewrite Eq.~(\ref{eq exact result}) as 
\begin{align}
 & \tilde\Sigma(t) = 	\\
 & \Delta S_X(t) - \sum_{x,e} \left[p_{xe}(t)\ln\bar p_{e|x}(\lambda_t) - p_{xe}(0)\ln\bar p_{e|x}(\lambda_0)\right]	\nonumber
\end{align}
where we used that the Shannon entropy of the global system $X\otimes E$ remains constant under Hamiltonian dynamics, 
$S_{XE}(t) = S_{XE}(0)$. Furthermore, we use the notation 
$\Delta f(t) \equiv f(t) - f(0)$ for any time-dependent function $f(t)$. Now, in accordance with phenomenological 
nonequilibrium thermodynamics, we would like to split $\tilde \Sigma(t)$ into two parts: 
\begin{equation}\label{eq 2nd law Seifert}
 \tilde\Sigma(t) = \Delta\tilde S_X(t) - \beta\tilde Q(t) \ge 0.
\end{equation}
Without additional information, there is obviously no unique splitting of these two quantities at this formal level, 
which essentially translates the results of Ref.~\cite{TalknerHaenggiPRE2016} into our framework. For the moment we will 
use the following definitions, which comply with the suggestions of Ref.~\cite{SeifertPRL2016}: 
\begin{align}
 \tilde S_X(t)	\equiv&~	S_X(t) + \lr{\beta^2\partial_\beta E_x^*(\lambda_t)}(t),	\label{eq def S Seifert}	\\
 \tilde Q(t)	\equiv&~	\beta^{-1} \sum_{x,e} \left[p_{xe}(t)\ln\bar p_{e|x}(\lambda_t) - p_{xe}(0)\ln\bar p_{e|x}(\lambda_0)\right]	\nonumber	\\
		&		+ \lr{\beta\partial_\beta E_x^*(\lambda_t)}(t) - \lr{\beta\partial_\beta E_x^*(\lambda_0)}(0).
\end{align}
The latter can be also rewritten as 
\begin{equation}
 \tilde Q(t) = -W(t) + \lr{\partial_\beta \beta E_x^*(\lambda_t)}(t) - \lr{\partial_\beta \beta E_x^*(\lambda_0)}(0),
\end{equation}
if we use the generally accepted definition for work $W(t) = \lr{E_{xe}(\lambda_t)}(t) - \lr{E_{xe}(\lambda_0)}(0)$. 
Assuming the first law of thermodynamics to be valid in the strong coupling case, this then implies a definition for 
internal energy:
\begin{equation}\label{eq def U Seifert}
 \tilde U_X(t) = \lr{\partial_\beta \beta E_x^*(\lambda_t)}(t) = \lr{E_x^* + \beta\partial_\beta E_x^*}(t).
\end{equation}
Introducing the non-equilibrium free energy 
\begin{equation}\label{eq def F Seifert}
 \tilde F_X(t) = \tilde U_X(t) - \frac{1}{\beta}\tilde S_X(t) = \lr{E_x^*(\lambda_t)}(t) - \frac{1}{\beta}S_X(t),
\end{equation}
we can alternatively write Eq.~(\ref{eq 2nd law Seifert}) as 
\begin{equation}\label{eq 2nd law Seifert 2}
 \tilde\Sigma(t) = \beta[W(t) - \Delta\tilde F_X(t)] \ge 0.
\end{equation}

The definitions above of $\tilde\Sigma$, $\tilde S_X$, $\tilde Q$, $\tilde U_X$ and $\tilde F_X$ seem to provide a 
satisfactory extension of thermodynamics to the strong-coupling case and they coincide with the definitions used by 
Seifert, who further motivates them by arguments of equilibrium statistical mechanics~\cite{SeifertPRL2016}. 
In addition to Ref.~\cite{SeifertPRL2016}, we have seen that the framework can be even extended by allowing for a 
time-dependence in the coupling $V_{xe}(t)$, too. 

Nevertheless, the approach above should be taken with care because it is ambiguous~\cite{TalknerHaenggiPRE2016}, 
and is not formulated at the level of instantaneous rates implying that the positivity of entropy 
production~(\ref{eq 2nd law Seifert}) crucially relies on the choice of initial state~(\ref{eq initial state}).

\subsection{Reduced thermodynamics description in $X\otimes Y$}
\label{sec HMF reduced}

We now clarify this situation by returning to our previous results in Sec~\ref{sec time-scale separation}. 
where we assumed that the environment is made of two parts, $E = Y \otimes R$. The first part $Y$ 
is strongly coupled to the system $X$ and is explicitly described. The second part $R$ is an ideal weakly coupled and 
Markovian thermal reservoir. Under these assumptions the ME~(\ref{eq ME}) and the full 
Hamiltonian dynamics give rise to the same description in the reduced space $X\otimes Y$. This implies, e.g., 
that the work computed within the ME framework [see Eq.~(\ref{eq work})] coincides with the work computed 
using the exact Hamiltonian dynamics as in Sec.~\ref{sec HMF exact}.

In the limit of TSS, 
$Y$ instantaneously equilibrates with respect to a given microstate of $X$. Thus, the initial 
requirement~(\ref{eq initial state}) is not only fulfilled initially but at any time $t$. 
This implicitly means that for any fixed value of $\lambda_t$, the global equilibrium steady state reads 
\begin{equation}
 p_{xyr} = \frac{e^{-\beta E_{xy}(\lambda_t)}}{\C Z_{XY}(\lambda_t)} \frac{e^{-\beta E_r}}{\C Z_R}
\end{equation}
where $E_r$ is the bare Hamiltonian of $R$. 
As a result, the HMF introduced in Eq.~(\ref{eq HMF standard definition}) \emph{coincides} with 
\begin{equation}\label{eq relation HMF free energies}
 E_x^* = E_x - \frac{1}{\beta}\ln\lr{e^{-\beta V_{xy}}}_Y^\text{eq} = F_{Y|x} - \C F_Y,
\end{equation}
which can be regarded as the HMF of $X \otimes Y$ only. 
Using the last equation together with~(\ref{eq cond U TSS}) and~(\ref{eq cond S TSS}), 
it is not hard to deduce the following two relations: 
\begin{align}
 \tilde U_X(t)	&=	U_{XY}(t) - \C U_Y,	\label{eq rel U}	\\
 \tilde S_X(t)	&=	S_{XY}(t) - \C S_Y.	\label{eq rel S}
\end{align}
Thus, apart from a time-independent additive constant the definitions for internal energy and system entropy 
\emph{within TSS coincide} with the definitions based on the HMF. Furthermore, since both approaches agree on the 
definition of work, we can show for the heat flow that 
\begin{equation}
 \dot{\tilde Q}(t) = d_t \tilde U_X(t) - \dot W(t)  = d_t U_{XY}(t) - \dot W(t) = \dot Q(t).
\end{equation}
Thus, within TSS we agree on this definition too and are able to derive the first law at the level of 
instantaneous rates. Likewise, we can also prove the positivity of the entropy production \emph{rate} by noting that 
\begin{equation}
 \begin{split}
  \dot{\tilde\Sigma}(t)	&=	d_t\tilde S_X(t) - \beta\dot{\tilde Q}(t)	\\
			&=	d_t S_{XY}(t) - \beta\dot Q(t) = \dot\Sigma(t) \ge 0.
 \end{split}
\end{equation}

As a preliminary summary we have thus shown that within TSS, the framework introduced in 
Ref.~\cite{SeifertPRL2016} is thermodynamically consistent and can be greatly extended. 
Furthermore, no ambiguity is left within our approach which allows us to refute the 
criticism raised in Ref.~\cite{TalknerHaenggiPRE2016} for our setup. 

It is interesting to ask what happens \emph{away from TSS} when $Y$ does not 
instantaneously conditionally equilibrate and $p_{y|x}$ is thus dynamically evolving. 
It is then possible to show that the framework of Sec.~\ref{sec HMF exact} does \emph{not} coincide with 
the original thermodynamic description of Sec.~\ref{sec setup} anymore. For instance, we prove 
in Appendix~\ref{sec Appendix} that the difference in entropy production can be expressed as 
\begin{equation}
 \begin{split}\label{eq diff entropy productions}
  \tilde\Sigma(t) - \Sigma(t)	&=	\beta(\Delta F_{XY} - \Delta\tilde F_X)	\\
				&=	D[p_{xy}(t)\|p_x(t)\bar p_{y|x}(\lambda_t)] \ge 0. 
 \end{split}
\end{equation}
Thus, $\tilde\Sigma(t)$ overestimates $\Sigma(t) = \int_0^t ds\dot\Sigma(s)$ by the relative entropy between 
the true state of $X\otimes Y$ and an idealized state of the form~(\ref{eq initial state}). 
Also, the rate of change of $\tilde\Sigma(t)$ can be negative. This and other features 
are explicitly demonstrated in the next section with the help of an example where the ME description 
in $X\otimes Y$ exactly coincides with the reduced Hamiltonian dynamics. 

To conclude, when it is possible to separate out the strongly coupled and non-Markovian degrees 
of freedom $Y$ from the environment $E$, then the following hierarchy of inequalities holds, 
\begin{equation}\label{eq 2nd law hierarchy}
 \tilde\Sigma(t) \ge \Sigma(t) \ge \Sigma^{(1)}(t) \ge 0.
\end{equation}
The equality $\tilde\Sigma(t) = \Sigma(t) = \Sigma^{(1)}(t)$ holds in the limit of TSS. Each of the entropy production 
in Eq.~(\ref{eq 2nd law hierarchy}) corresponds to a different layer of the description. $\Sigma^{(1)}(t)$ assumes 
$Y$ to be conditionally (versus $X$) equilibrated and, of course, implicitly $R$ to be equilibrated. $\Sigma(t)$ assumes 
only an ideal reservoir $R$ and $\tilde\Sigma(t)$ is an exact result which can be applied to any Hamiltonian dynamics 
(\ref{Tot_Hamiltonian}) as long as the initial condition is of the form (\ref{eq initial state}).

\section{Discrepancy in the non-Markovian regime}
\label{sec example}

Within TSS, i.e., whenever the environment behaves Markovian by instantaneously adapting to the microstate of the 
system $X$, we have proven the equivalence of the coarse-grained thermodynamic framework from 
Sec.~\ref{sec time-scale separation} with the approach based on the HMF. In principle, both frameworks can be also
applied beyond TSS and we will now provide a counterexample showing that the HMF-approach then no longer coincides 
with the standard framework of Sec.~\ref{sec setup}. 

We consider the example of driven Brownian motion thereby demonstrating that our main results above do not only hold for 
dynamics on discrete states but also for continuous variables. The global Hamiltonian with mass-weighted coordinates 
is specified by~\cite{WeissBook2008, SekimotoBook2010} 
\begin{align}\label{eq Brownian motion Hamiltonian}
 E_{xe}(t)	&=	E_x(t) + V_{xe} + E_e,	\\
 E_x(t)		&=	\frac{1}{2}[p_x^2 + \omega^2(t)x^2],	\\
 V_{xe} + E_e	&=	\frac{1}{2}\sum_k\left[p_k^2 + \nu_k^2\left(x_k - \frac{c_k}{\nu_k^2}x\right)^2\right]
\end{align}
and we identify $\lambda_t = t$ in the following and use $\omega(t) = \omega_0 + g\sin(\omega_Lt)$. 
We relaxed the notation meaning with $E_x(t)$ the energy associated to the microstate $(x,p_x)$ and 
a microstate $e$ of the bath is given by specifying $(x_k,p_k)$ for all $k$. 
Furthermore, the spectral density (SD) of the bath is defined as and parametrized by 
\begin{equation}\label{eq SD non-Markovian}
 J(\omega) \equiv \frac{\pi}{2}\sum_k\frac{c_k^2}{\nu_k}\delta(\omega-\nu_k) = \frac{\lambda_0^2\gamma\omega}{(\omega^2-\omega_1^2)^2 + \gamma^2\omega^2}.
\end{equation}
Here, $\lambda_0$ controls the overall coupling strength between the system and the environment and $\gamma$ changes the 
shape of the SD from a pronounced peak around $\omega_1$ for small $\gamma$ to a rather unstructured and flat SD for 
large $\gamma$. 

The corresponding Langevin equation for this setup reads~\cite{WeissBook2008, SekimotoBook2010} 
\begin{equation}\label{eq Langevin eq general}
 \ddot x(t) + \omega_0^2(t) x(t) + \int_0^t ds \Gamma(t-s)\dot x(s) = \xi(t)
\end{equation}
with the friction kernel 
\begin{equation}
 \Gamma(t) \equiv \int_0^\infty d\omega \frac{2}{\pi\omega} J(\omega) \cos(\omega t)
\end{equation}
and the noise $\xi(t)$, which obeys the statistics 
\begin{equation}
 \lr{\xi(t)}_E = 0, ~~~ \lr{\xi(t)\xi(s)}_E = \frac{1}{\beta}\Gamma(t-s).
\end{equation}
We see that for an Ohmic SD $J(\omega) = \eta\omega$ (times a high-frequency cutoff as usual) we obtain 
$\Gamma(t) = 2\eta\delta(t)$ and this gives the standard Langevin equation with Gaussian white noise. 
Unfortunately, our SD~(\ref{eq SD non-Markovian}) is not Ohmic unless we scale 
$\lambda_0 = \sqrt{\alpha_1\alpha_2\gamma}$, $\omega_1 = \sqrt{\alpha_2\gamma}$ and send $\gamma\rightarrow\infty$. 
This implies an Ohmic SD for sufficiently large $\alpha_2$: 
\begin{equation}
 \lim_{\gamma\rightarrow\infty} J(\omega) = \alpha_1\frac{\alpha_2\omega}{\alpha_2^2 + \omega^2}.
\end{equation}

Establishing a consistent thermodynamic framework for the general Langevin Eq.~(\ref{eq Langevin eq general}) cannot be 
done using standard tools from stochastic thermodynamics. One route, however, could be to take the definitions from 
Sec.~\ref{sec connection HMF} and to apply them here. Application of these definitions is facilitated by the fact that 
for a Brownian motion Hamiltonian the HMF \emph{coincides} with the bare system Hamiltonian, i.e., $E_x^* = E_x$, which 
can be directly checked by evaluating the Gaussian integrals. 
Thus, the change in internal energy and system entropy read $\Delta\tilde U_X = \lr{E_x(t)}(t) - \lr{E_x(0)}(0)$ and 
$\Delta\tilde S_X = \lr{-\ln p_x(t)}(t) - \lr{-\ln p_x(0)}(0)$. That is to say, the HMF-approach uses for this examples 
the \emph{standard} weak-coupling definitions \emph{irrespective} of the spectral properties of the bath. 
Furthermore, since work can be computed using $W(t) = \int_0^t ds\lr{d_t E_x(s)}(s)$, we obtain $\tilde Q$ and 
$\tilde\Sigma$, too. However, to access the dynamics of the system, we would have to simulate the non-Markovian 
Langevin equation~(\ref{eq Langevin eq general}), which is numerically demanding. 

We therefore follow a different strategy and identify a subsystem $Y\subset E$, which transforms the non-Markovian 
system $X$ to a Markovian system $X\otimes Y$. This is most conveniently done by identifying a collective degree of 
freedom in the environment defined via 
\begin{equation}
 \lambda_0 y \equiv \sum_k c_k x_k.
\end{equation}
In this context, $y$ is also known as a reaction coordinate. It has been shown to successfully model the 
dynamics of non-Markovian open quantum systems (see, e.g., Refs.~\cite{PriorEtAlPRL2010, MartinazzoEtAlJCP2011, 
WoodsEtAlJMP2014, IlesSmithLambertNazirPRA2014}) and has been recently proposed as a method 
to establish a consistent thermodynamic framework beyond the Markovian and weak-coupling 
approximation~\cite{StrasbergEtAlNJP2016, NewmanMintertNazirPRE2017}. 

We skip the details of the derivation, which can be looked up in the literature~\cite{PriorEtAlPRL2010, 
MartinazzoEtAlJCP2011, WoodsEtAlJMP2014, IlesSmithLambertNazirPRA2014, StrasbergEtAlNJP2016, NewmanMintertNazirPRE2017}, 
and only state the main result. After the transformation, the Hamiltonian becomes 
\begin{align}
 E_{xyr}(t)	&=	E_x(t) + V_{xy} + E_y + V_{yr} + E_r,	\\
 V_{xy} + E_y	&=	\frac{\lambda_0^2}{2\omega_1^2}x^2 - \lambda_0 x y + \frac{1}{2}(p_y^2 + \omega_1^2 y^2),	\\
 V_{yr} + E_r	&=	\frac{1}{2}\sum_k\left[\tilde p_k^2 + \tilde \nu_k^2\left(\tilde x_k - \frac{\tilde c_k}{\tilde\nu_k^2}y\right)^2\right],
\end{align}
where the new SD of the ``residual environment'' $R$ is defined and for the choice~(\ref{eq SD non-Markovian}) given by 
\begin{equation}
 \tilde J(\omega) \equiv \frac{\pi}{2}\sum_k\frac{\tilde c_k^2}{\tilde \nu_k}\delta(\omega-\tilde \nu_k) = \gamma\omega.
\end{equation}
This SD immediately yields the coupled set of \emph{Markovian} Langevin equations 
\begin{equation}
 \begin{split}\label{eq Langevin eq RC}
  \ddot x(t) + \left[\omega_0^2(t) + \frac{\lambda_0^2}{\omega_1^2}\right] x(t) - \lambda_0 y(t)	&=	0,	\\
  \ddot y(t) + \gamma\dot y(t) + \omega_1^2 y(t) - \lambda_0 x(t)	&=	\xi(t)
 \end{split}
\end{equation}
with Gaussian white noise $\xi(t)$. 

Following standard procedures~\cite{SekimotoBook2010}, we can associate a Fokker-Planck equation for the probability 
distribution $P(t) = P(x,p_x,y,p_y;t)$ to the set of Langevin equations above. It reads 
\begin{equation}\label{eq FPE}
 \partial_t P(t) = \left(-\nabla^T\cdot A\cdot \bb x + \frac{1}{2}\nabla^T\cdot B\cdot\nabla\right)P(t).
\end{equation}
where we defined $\nabla \equiv (\partial_{x},\partial_{p_x},\partial_{y},\partial_{p_y})^T$, 
${\bb x \equiv (x,p_x,y,p_y)^T}$ and introduced the matrices 
\begin{equation}
 A = \left(\begin{array}{cccc}
            0	&	1	&	0	&	0	\\
            -[\omega_0^2(\lambda_t) + \lambda_0^2/\omega_1^2]	&	0	&	\lambda_0	&	0	\\
            0	&	0	&	0	&	1	\\
            \lambda_0	&	0	&	-\omega_1^2	&	-\gamma	\\
           \end{array}\right)
\end{equation}
and $B$ whose only non-zero component is $B_{44} = 2\gamma/\beta$. 
We emphasize that Eq.~(\ref{eq FPE}) describes the exact dynamics in $X\otimes Y$. No approximation 
has been made in any of the steps above (apart from assuming an initially equilibrated reservoir state). 

An advantage of this Fokker-Planck equation is that the dynamics of the first and second cumulants are closed. 
In fact, the equations of motion for the first cumulants $\lr{z}$ (with $z\in\{x,p_x,y,p_y\}$) couple only to 
themselves and the same is true for the second cumulants $C_{zz'} \equiv \lr{zz'} - \lr{z}\lr{z'}$. Thus, an 
initially Gaussian state will stay Gaussian for all times. Computing the time-evolution of the first two cumulants 
based on an initial condition of the form~(\ref{eq initial state}) can then be easily done numerically. 
Because standard stochastic thermodynamics applies to Eqs.~(\ref{eq Langevin eq RC}) or~(\ref{eq FPE}), we 
have direct access to averaged thermodynamic quantities for $X\otimes Y$ introduced in Sec.~\ref{sec setup}, 
also see Ref.~\cite{SekimotoBook2010}. Furthermore, because we have the exact dynamics in 
$X\otimes Y$, we also get the exact reduced dynamics of $X$ by tracing over $Y$, consequently giving direct access 
to the time evolution of $\tilde U_X,\tilde S_X,\tilde Q$ and $\tilde\Sigma$. Thus, our Markovian embedding strategy 
has allowed us to circumvent the difficulty to simulate the non-Markovian Langevin equation~(\ref{eq Langevin eq general}). 

\begin{figure*}
 \centering\includegraphics[width=0.99\textwidth,clip=true]{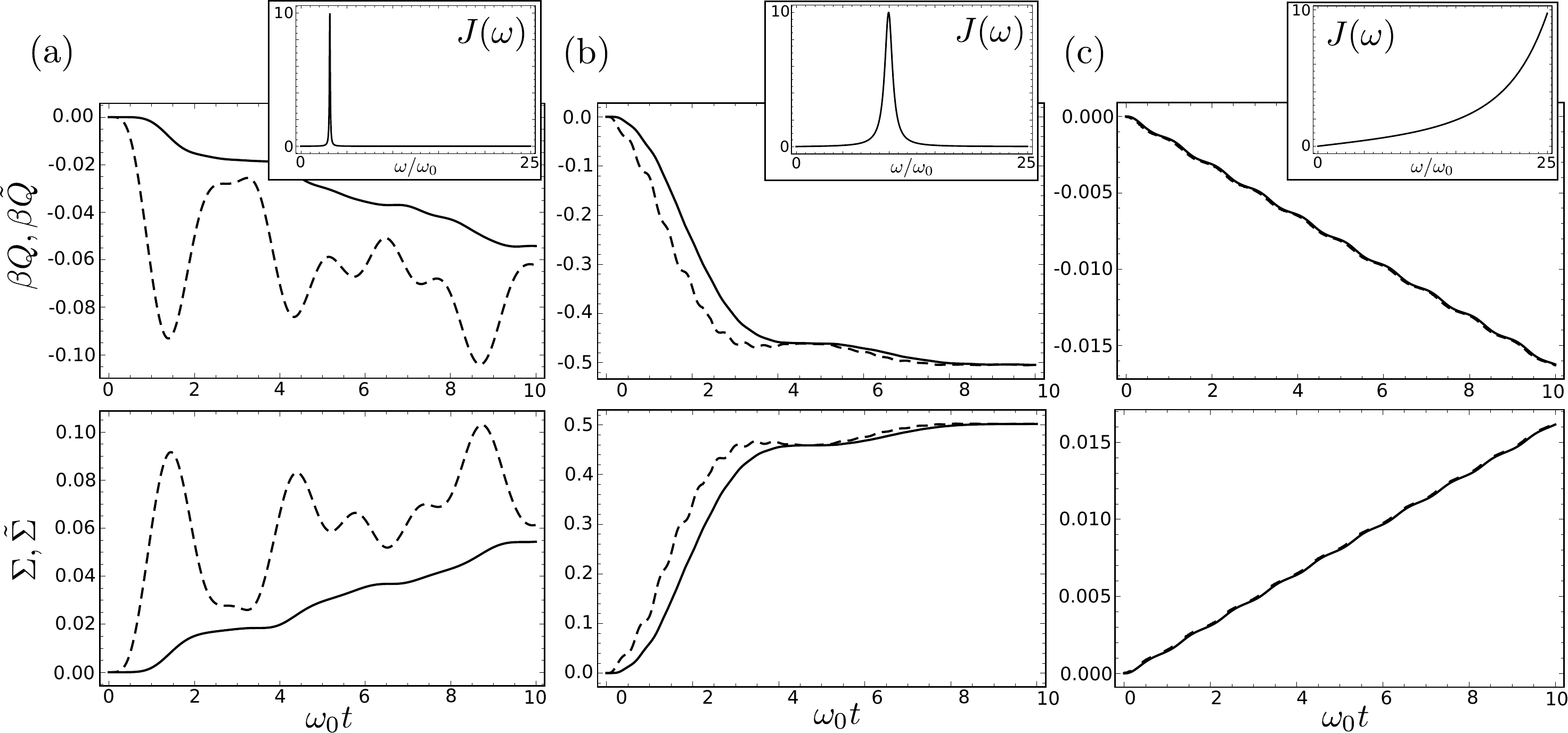}
 \label{fig thermo Brownian motion} 
 \caption{Plot of the thermodynamics for a driven Brownian particle coupled to a non-Markovian environment. 
 Each column (a), (b), (c) refers to a different form of the SD displayed in the upper right corner of each column. 
 Below the SD, we show two different plots: the upper one displays 
 $Q(t)$ (solid line) and $\tilde Q(t)$ (dashed line) as a function of dimensionless time $\omega_0t$; the lower 
 one displays $\Sigma(t)$ (solid line) and $\tilde\Sigma(t)$ (dashed line). Parameters for the driving are 
 $\omega_L = 2\pi\omega_0$ and $g = 0.1\omega_0$. The SD is parametrized as $\lambda_0 = \sqrt{\alpha_1\alpha_2\gamma}$ 
 and $\omega_1 = \sqrt{\alpha_2\gamma}$ with $\alpha_2 = 100\omega_0$. The SDs differ by the choice of 
 $(\gamma,\alpha_1)$, which we chose as $(0.1\omega_0,\omega_0^2)$ in (a), $(\omega_0,10^2\omega_0^2)$ in (b) 
 and $(10\omega_0,250^2\omega_0^2)$ in (c). The system was prepared using Eq.~(\ref{eq initial state}) and had 
 initial mean values $\lr{x}(0) = (\sqrt{\beta}\omega_0)^{-1}$, $\lr{p_x}(0) = 0$ and covariances 
 $C_{xx}(0) = (\beta\omega_0^2)^{-1}, C_{p_xp_x}(0) = \beta^{-1}$ and $C_{xp_x}(0) = 0$. 
 Finally, we set $\omega_0 = 1$ and $\beta = 1$. }
\end{figure*}

Results of the simulation are shown in Fig.~\ref{fig thermo Brownian motion}. We vary the SD from a strongly 
non-Markovian situation (shown on the left) to a Markovian but strong-coupling situation (on the right) by changing 
$\gamma$ and $\alpha_1$. For each $\gamma$ we compare the integrated heat flows $Q$ and $\tilde Q$ (upper panel) and the integrated 
entropy production $\Sigma$ and $\tilde\Sigma$ (lower panel). The following main features are observable: 
for large $\gamma$ the assumption of TSS is justified and quantities defined in 
Sec.~\ref{sec time-scale separation} and~\ref{sec connection HMF} agree perfectly. 
In fact, $\gamma$ is directly linked to the rate of relaxation of the reaction coordinate $(y,p_y)$, but does not 
directly couple to the system degrees of freedom $(x,p_x)$. Thus, a large $\gamma$ corresponds to the limit 
of TSS as introduced in Eq.~(\ref{eq TSS limit}). Away from that limit, however, 
we observe that $\tilde Q$ differs \emph{significantly} from $Q$ and the same observation is true for the different 
definitions of entropy production, too. Also, although Eq.~(\ref{eq 2nd law Seifert}) is always 
obeyed, the rate of $\tilde\Sigma$ can become negative. Furthermore, we can also confirm the validity of 
Eq.~(\ref{eq diff entropy productions}), $\tilde\Sigma(t)\ge\Sigma(t)$.

\section{Summary}
\label{sec summary}

We clarified important questions in the framework of strong-coupling thermodynamics. 
Our main achievements are the following: 

\emph{1) Justification of the HMF within TSS.} Within the limit of TSS, the framework provided in 
Ref.~\cite{SeifertPRL2016} is thermodynamically consistent for arbitrary system states $p_x(t)$ 
and the HMF is a legitimate tool to investigate the thermodynamics of systems in strong contact with a 
\emph{single} environment. 

\emph{2) No ambiguity. } Any ambiguity is removed in our framework and the concerns put forward in 
Ref.~\cite{TalknerHaenggiPRE2016} 
do not apply. The reason for this is that we start from a well-defined weak coupling framework. Especially and 
contrary to previous attempts, we do not use the first law to define heat, but have an alternative and 
unambiguous definition for it. 

\emph{3) Extension of previous results. } Thanks to the TSS, 
we were able to significantly extend previous results by formulating them at the level of instantaneous rates 
instead of integrated quantities and by allowing also for a time-dependence in the system-environment coupling. 

\emph{4) Difficulties in the non-Markovian regime. }
Away from TSS, the framework of Ref.~\cite{SeifertPRL2016} does not match the original thermodynamic 
picture though Eq.~(\ref{eq exact result}) is always obeyed. Thus, we observe that in order to establish the original laws 
of thermodynamics in the non-Markovian regime (where the environment is also dynamically evolving), one is forced to 
fully take into account the (thermo)dynamics of $X$ \emph{and} $Y$. Any effective description at this stage will in 
general miss important pieces in the first or second law. This complies with the point of view put forward in 
Ref.~\cite{StrasbergEtAlNJP2016, NewmanMintertNazirPRE2017}. 

Recently, two alternative approaches were put forward in Ref.~\cite{JarzynskiPRX2017} by starting from the 
isothermal-isobaric ensemble and by taking pressure and volume effects into account. These approaches correctly 
reproduce the macroscopic limit by introducing the notion of ``thermodynamic volume'' for a microscopic system. 
The ``bare representation'' in Ref.~\cite{JarzynskiPRX2017} shows that it is possible to retain the original weak 
coupling definitions of internal energy and entropy by shifting our attention to enthalphy and Gibbs free energy instead. 
If the isobaric $PV$-contribution is absent or blindly ignored, then the 
``partial molar representation'' in Ref.~\cite{JarzynskiPRX2017} coincides with the approach in 
Sec.~\ref{sec connection HMF}.\footnote{To compare notation, we have without $PV$-terms that the Gibbs free energies 
in Ref.~\cite{JarzynskiPRX2017} are related to our free energies via $G_0^{\C E} = \C F_E$,  
$G_x^{\C E} = F_{E|x} - E_x$ and the solvation Hamiltonian of mean force becomes $\phi(x) = E_x^* - E_x$. 
However, note that the $PV$-term in Ref.~\cite{JarzynskiPRX2017} is actually only negligible at weak coupling. 
Then, $E_x^* \approx E_x$ and all the different frameworks coincide with the weak coupling limit. } 
These alternative approaches should be therefore also derivable 
within TSS, but we expect that outside the limit of TSS they will mismatch again. 


Finally, we would like to mention that the framework of Sec.~\ref{sec connection HMF} can be used in principle also beyond 
TSS, for instance, if it is impossible to find a splitting $E = Y \otimes R$ or if the dynamical simulation of the 
environment becomes unfeasible. It then nevertheless has to be treated with care and further consistency checks still 
need to be carried out such as, for instance, the implication of the correct thermodynamic laws in the limit of 
reversible transformations as investigated in Ref.~\cite{EspositoOchoaGalperinPRB2015}. 

\emph{Acknowledgements. }
Financial support by the National Research Fund Luxembourg (project FNR/A11/02) 
and by the European Research Council (project 681456) is acknowledged.


\bibliography{books,open_systems,thermo,info_thermo,general_refs}

\appendix

\section{Relation between the entropy production in Eq.~(\ref{eq exact result}) and Ref.~\cite{EspositoLindenbergVandenBroeckNJP2010}}
\label{sec Appendix NJP}

In Ref.~\cite{EspositoLindenbergVandenBroeckNJP2010}, the entropy production of an Hamiltonian dynamics 
(\ref{Tot_Hamiltonian}) with an initial condition of the form $p_x(0)p_e^\text{eq}$
was defined as 
\begin{equation}
 \Sigma_\text{NJP}(t) \equiv D(p_{xe}(t)\|p_x(t)p_e^\text{eq}) \ge 0.
\end{equation}
This result is very close in spirit to the entropy production (\ref{eq exact result}) that 
we derived in this paper for the same Hamiltonian dynamics but with an initial condition 
of the form (\ref{eq initial state}).  
It measures the deviation of the true state from an idealized product state where 
the environment is always at equilibrium instead of conditionally at equilibrium. 

The only meaningful comparison between the two expressions requires to consider situations where the 
two classes of initial conditions coincide, namely when $V_{xe}(\lambda_0) = 0$ and the interaction 
is only turned on afterwards. 
In this case, we find that
\begin{align}
 & \tilde\Sigma(t) - \tilde\Sigma_\text{NJP}(t) = \sum_{x,e} p_{xe}(t)\ln\frac{p_e^\text{eq}}{\bar p_{e|x}(\lambda_t)}	\\
 & = \beta \sum_{x,e} p_{xe}(t) \left[V_{xe}(\lambda_t) - \left(-\frac{1}{\beta}\ln\lr{e^{-\beta V_{xe}(\lambda_t)}}_E^\text{eq}\right)\right].	\nonumber
\end{align}
This relation can be rewritten as a difference between the non-equilibrium free energy (\ref{eq def F Seifert})
and the non-equilibrium free energy corresponding to the scheme of Ref.~\cite{EspositoLindenbergVandenBroeckNJP2010}
\begin{equation}
 F_\text{NJP}(t) \equiv \lr{E_x(\lambda_t) + V_{xe}(\lambda_t)}(t) - \beta^{-1}S_X(t).
\end{equation}
Explicitly, 
\begin{equation}\label{eq diff 2nd law Seifert NJP}
 \tilde\Sigma(t) - \tilde\Sigma_\text{NJP}(t) = \tilde F_X(t) - F_\text{NJP}(t).
\end{equation}
In general, there is no bound for this difference. 

\section{Proof of Eq.~(\ref{eq diff entropy productions})}
\label{sec Appendix}

From Eq.~(\ref{eq entropy production alternative 2}) we deduce that $\Sigma(t) = \beta[W(t) - \Delta F_{XY}]$. 
Thus, with Eq.~(\ref{eq 2nd law Seifert 2}) we immediately get the first line of Eq.~(\ref{eq diff entropy productions}), 
\begin{equation}
 \tilde\Sigma(t) - \Sigma(t) = \beta(\Delta F_{XY} - \Delta\tilde F_X).
\end{equation}
To prove the second, we look at $F_{XY}(t) - \tilde F_X(t)$ and $F_{XY}(0) - \tilde F_X(0)$ in detail. Since the formalism 
using the HMF in Sec.~\ref{sec connection HMF} assumes that the environment starts in a conditionally equilibrated 
state $p_{xy}(0) = p_x(0)\bar p_{y|x}(\lambda_0)$, Eqs.~(\ref{eq rel U}) and~(\ref{eq rel S}) are valid at $t=0$. 
Straightforward algebra then gives 
\begin{equation}
 F_{XY}(0) - \tilde F_X(0) = \C F_Y.
\end{equation}
At later times, using the definitions~(\ref{eq def S Seifert}) and~(\ref{eq def U Seifert}), we find that 
\begin{equation}
 \begin{split}
  & F_{XY}(t) - \tilde F_X(t)	\\
  & = \lr{E_{xy}(\lambda_t)} - TS_{XY}(t) - \lr{E_x^*(\lambda_t)} + TS_X(t). 
 \end{split}
\end{equation}
Next, from Eq.~(\ref{eq relation HMF free energies}) together with~(\ref{eq cond equilibrium Y}) we obtain 
\begin{equation}
 \lr{E_x^*(\lambda_t)} = \lr{E_{xy}(\lambda_t)} + \beta^{-1}\lr{\ln \bar p_{y|x}(\lambda_t)} - \C F_Y.
\end{equation}
Thus, we have explicitly 
\begin{equation}
 F_{XY}(t) - \tilde F_X(t) = \C F_Y + \beta^{-1}\sum_{x,y} p_{xy}(t)\ln\frac{p_{y|x}(t)}{\bar p_{y|x}(\lambda_t)}
\end{equation}
and consequently, 
\begin{equation}
 \beta(\Delta F_{XY} - \Delta\tilde F_X) = \sum_{x,y} p_{xy}(t)\ln\frac{p_{y|x}(t)}{\bar p_{y|x}(\lambda_t)},
\end{equation}
which proves the second line of Eq.~(\ref{eq diff entropy productions}).

\end{document}